\documentstyle[ppmo209,twoside,epsf,fancyhdr]{article}
\input{psfig.sty}                               
\newcommand{\lsim}{\, \lower2truept\hbox{${< \atop\hbox{\raise4truept\hbox{$\sim$}}}$}\,}
\newcommand{\gsim}{\, \lower2truept\hbox{${> \atop\hbox{\raise4truept\hbox{$\sim$}}}$}\,}
\newcommand{\nid}{\noindent}

\begin{document}

\markboth {} {Stephan's Quintet: A Multi-galaxy Collision}

\title{Stephan's Quintet: A Multi-galaxy Collision}

\author{C. Kevin Xu$^{1}$\mailto{}}

\inst{$^1$ Infrared Processing and Analysis Center, 
California Institute of Technology 100-22, Pasadena, CA 91125, USA}

\email{cxu@ipac.caltech.edu}


\begin{abstract}
Stephan's Quintet (SQ), discovered more than 100 years ago, is the
most famous and well studied compact galaxy group.  It has been
observed in almost all wavebands, with the most advanced instruments
including Spitzer, GALEX, HST, Chandra, VLA, and various large
mm/submm telescopes/arrays such as the IRAM 30m and BIMA. The rich
multi-band data reveal one of the most fascinating pictures in
the universe, depicting a very complex web of interactions between
member galaxies and various constituents of the intragroup medium
(IGM), which in turn trigger some spectacular activities such as a
40 kpc large scale shock and a strong IGM starburst. In this talk I
will give a review on these observations.


\end{abstract}


\setlength\baselineskip {5mm}             


%
%

\vskip1cm
\section{Introduction}           
\label{sect:intro}

As an interacting galaxy system, Stephan's Quintet (hereafter SQ) is
not as famous as Arp~220 or M82. Taken as a 'popularity index', the
number of references listed by NED (until Dec. 18th, 2005) under
'Stephan's Quintet' is only 104, far less than that under 'M82' (1355)
and 'Arp~220' (563).  However, its fame as a very special case of
multi-galaxy collision has been rising steadily recently, thanks
mostly to the ever increasing number of observational windows opened
one after another in this space astronomy era.  SQ has been observed
in almost all wavebands, and it keeps revealing
surprises when being looked at by new instruments.

Different from binary galaxies and mergers which have been thoroughly
discussed elsewhere in this conference, SQ belongs to a different
class of interacting systems of galaxies: the compact
groups (Hickson~\cite{hic82}) that are characterized by aggregates of 4 -- 8
galaxies in implied space densities as high as those in cluster cores.
SQ is unique among its peers because a seemingly 
very rare event is occurring to it: a high velocity ($\sim
1000$ km sec$^{-1}$) intruder is colliding into a debris field in
the intragroup medium (IGM), the latter being a product of previous
interactions among other member galaxies of the group.
The complex web of galaxy-galaxy and galaxy-IGM interactions
have triggered all kinds of interaction related phenomena,
including a large scale shock ($\sim 40$ kpc), an IGM starburst,
long tidal tails ($> 100$ kpc) with tidal dwarf candidates,
and a highly obscured type II AGN.

In this talk I'll review the rich literature 
on the multi-frequency observations of SQ and the
constraints imposed on its interaction history 
and on its current star formation rate. I'll also
high light its
role as a laboratory to study rare phenomena such as
large scale shocks and high velocity collision induced 
starbursts. Inferences from these studies to the
understanding of local ULIRGs and high z galaxies
will be briefly touched.

\section{Early Literature and Membership of SQ}
\begin{figure}
   \vspace{2mm}
   \begin{center}
   \centering
    \hspace{3mm}\psfig{figure=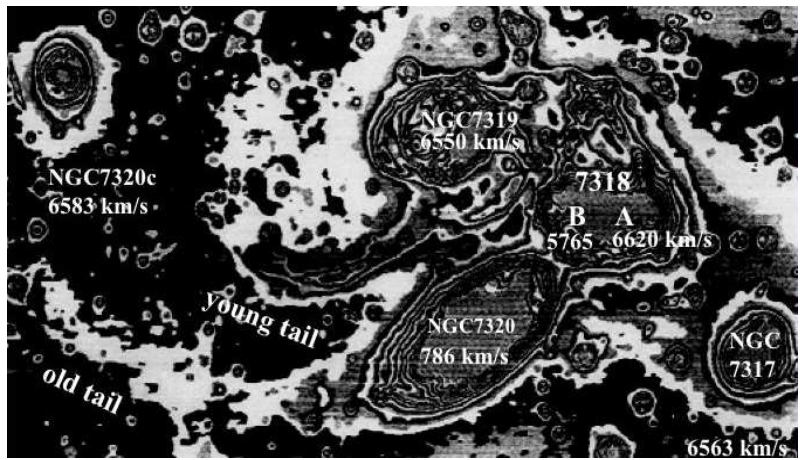,width=105mm,height=60mm,angle=0.0}
   \caption{SQ members and two tidal tails. The isodensity image of
a photographic plate is taken from Arp (\cite{arp73a}).}
   \label{Fig:lightcurve-ADAri}
   \end{center}
\end{figure}
SQ (Fig.1) was discovered as an aggregation of nebulae
by Stephan in late 19th century (Stephan~\cite{ste1877}), 
much earlier than 
anyone knew galaxies outside the Milky Way exist.
It raised strong interest
in early years of modern astronomy because of
its wide range of redshifts
(Ambartsumian~\cite{amba58}; Limber \& Mathews~\cite{limb60};
Burbidge \& Burbidge~\cite{bb71}).
In particular, Arp(\cite{arp73a},\cite{al76}) argued that NGC~7320, with a redshift 
about 5000 km sec$^{-1}$ less than the rest of the group
(Burbidge \& Burbidge~\cite{bb71}), is physically associated with
other SQ galaxies and is a prime case 
for the existence of non-Dopler redshift. 
The counter arguments against this claim include (Allen 
 \& Sullivan~\cite{allen80}; Moles et al.~\cite{mol97}): 
(1) No evidence
for interaction with other SQ galaxies was found in
the HI or in the H$\alpha$ emission associated with NGC~7320. 
(2) Results from redshift independent
estimators yielded distances consistent with
the redshifts. (3) Apparently NGC~7320 is a member
of a foreground loose galaxy group which includes the large galaxy NGC~7331
at redshift $\sim 800$ km sec$^{-1}$. I shall assume that
NGC~7320 is a foreground galaxy and omit it hereafter.
Arp (\cite{arp73b}) 
suggested that NGC~7320c,  a galaxy about 4' to the east of NGC~7319
and linked to the group by optical tidal tails,
should be included into SQ. Therefore,
dropping NGC~7320 and at same time
picking up NGC~7320c, SQ still consists of five galaxies as
implied by the name.

\section{Interaction History}
The early deep optical images (Arp \& Kormendy~\cite{ak72}; 
Arp~\cite{arp73a}; Arp \& Lorre~\cite{al76}) show clearly two concentrating
parallel tidal tails, both starting from NGC~7319
and extending to NGC~7320c, with the more diffuse one
passing through the low redshift galaxy NGC~7320.
However, contemplations
on the interaction history of SQ started only after two discoveries in the
radio band, both made with the Westerbork Synthesis Telescope.
The first is a huge ($\sim 40$ kpc)
radio continuum ridge found in the IGM
between NGC~7319 and NGC~7318b (Allen \& Hartsuiker~\cite{allen72}). 
Among other possible interpretations,
Allen \& Hartsuiker suggested this may 
be a large scale shock triggered
by an on-going collision between NGC~7318b and the
rest of the group. The second (Allen \& Sullivan~\cite{allen80}; 
Shostak et al.~\cite{sho84}) is the revelation that nearly
all HI gas associated with SQ, amounting to $\sim 10^{10}$
M$_{\sun}$, is in the IGM outside galaxy disks.
It is suggested that this is stripped gas from
late type galaxy disks due to an earlier interaction (a few 
10$^8$ yr ago) either between NGC~7319 and NGC~7318a 
(Allen \& Sullivan~\cite{allen80}), or between NGC~7319 and NGC~7320c
(Shostak et al.~\cite{sho84}).

Mole et al. (\cite{mol97}, hereafter M97) put up a comprehensive picture
for the interaction history of SQ based on
a 'two-intruders' scenario (see also Sulentic et al.~\cite{s01}, hereafter S01): 
an old
intruder (NGC~7320c) stripped most of the gas from group members, and
a new intruder (NGC~7318b) is currently colliding with this gas and
triggering the large scale shock. The 'old intruder'
passed the core of SQ twice, one in $\sim 5$ -- 7 $10^8$ yr ago
and pulled out the `old tail', another
about 2 $10^8$ yr ago and triggered the `young tail' (the narrower one).
Recently, Xu et al. (\cite{xu05}, hereafter X05) proposed a 'three-intruders' scenario
based on new observations. They agreed with M97 on the origins of the old tail
and of the shock. But, different from M97, they argued that it is
unlikely that the 'young tail' is also triggered by NGC~7320c. 
This is because the
recently measured redshift of NGC~7320c (6583 km sec$^{-1}$, S01)
is almost identical to that of
NGC~7319, indicating a slow passage (S01) rather
than a fast passage ($\sim 700$ km sec$^{-1}$, M97). 
In order for NGC~7320c to move to its current position, the
NGC~7319/7320c encounter must have occurred $\gsim 5\; 10^8$ yr ago.
This is close to the age of the old tail, but older than that of the
young tail. They suggested that the young tail is
triggered by a close encounter between the elliptical galaxy
NGC~7318a and NGC~7319. The projected
distance between NGC~7318a/7319 is only $\sim 1/3$ of that between
NGC~7320c/7319. Therefore the time argument is in favor of the new scenario.
Also the morphology of the 'young tail' revealed by the new UV
observations, relative to NGC~7319 and NGC~7318a, 
looks very similar to the 'counter tidal tail' found in 
dynamic simulations of equal mass galaxy-galaxy interactions 
(see, e.g., Fig.2 of Toomre \& Toomre~\cite{tt72}, t=2 10$^8$ yr).
According to the K-band luminosities,  NGC~7318a and NGC~7319
have nearly identical stellar mass.

Moles et al. (\cite{mol97}; \cite{mol98}) concluded that the disk of NGC~7318b is
still intact because the time since it starts to interact with the
group is too short ($\sim 10^7$ yr) for any tidal effect. However,
Xu et al. (\cite{xu03}; \cite{xu05}) speculated that the galaxy might have had
a head-on collision with NGC~7318a $\sim 10^8$ yr
ago, which means NGC~7318a must be $\sim 100$ kpc further away than
NGC~7318b given the velocity difference. This will help to explain the
long outer arms (looking like tidal tails), the peculiar HI gas
distribution (all outside the optical disk) and the huge UV disk ($\sim
80$ kpc) of the NGC~7318b. Also, in this scenario,
the comma-like filaments in the region north of NGC7318 
(see, e.g. Fig.1c of Mendes de Olivera et al.~\cite{mdo01}) can be interpreted
in an analogy to the spoke-like filaments in the Cartwheel Galaxy,
a prototype ring galaxy produced in a head-on galaxy-galaxy collision 
(Higdon~\cite{hig95}).

The only sign for any interaction between NGC~7317 with other members
of SQ is a common halo linking it to the binary NGC~7318,
found in the deep X-ray and optical R-band images (Trinchieri
et al.~\cite{trin05}).

\section{The Large Scale Shock}
Since the discovery of Allen \& Hartsuiker (\cite{allen80}),
the large scale radio ridge has been confirmed by later VLA observations 
(van der Hulst \& Rots~\cite{van81}; Williams et al.~\cite{w02}, hereafter W02;
Xu et al.~\cite{xu03}). The high resolution X-ray
maps (Pietsch et al.~\cite{pie97}; S01;
Trinchieri et al.~\cite{trin03}, \cite{trin05}) and optical narrow-band
H$\alpha$/[NII] maps (Vilchez \&
Iglesias-P\'aramo~\cite{vi98}; Ohyama et al.~\cite{oh98};
Xu et al.~\cite{xu99}; Plana et al.~\cite{plana99}; S01)
show a similar structure, in both the
hot gas and the ionized gas.
The optical spectroscopy of Xu et al. (\cite{xu03}) at several 
positions along the ridge show spectra consistent with shock
excitation.
This shock is truly unique to SQ. 
Its size is second only to that of the radio relics caused
by cluster mergers (En{\ss}lin \& Br\"uggen~\cite{en02}). 
In the shock, the intruder/IGM
collision could have deposited as much as 10$^{56}$ erg
energy, equivalent to the kinematic energy of 
$\sim 100,000$ supernovae.
Using the formula of Dopita \& Sutherland (\cite{dopita96}), 
Xu et al. (\cite{xu03}) derived a total energy flux
in the form of $L_{\rm T} =  2.1\times10^{42}\,\,{\rm
erg~sec}^{-1}$.
Apparently, most of this flux comes out in the FIR band. 
Based on ISO observations,
the total dust luminosity of the shock front is
$L_{{\rm dust}}=1.9~10^{42} {\rm erg~s}^{-1}$. This is about an order
of magnitude higher than the X-ray luminosity (Trinchieri
et al.~\cite{trin03}), indicating that the hot gas is predominantly
cooled by collisions with dust grains.
Trinchieri
et al. (\cite{trin03}) estimated the shock velocity to be 460 km sec$^{-1}$,
and the density of post-shock gas $n_{\rm H} = 0.027$
cm$^{-3}$. The corresponding sputtering time scale
for $a=0.1\mu m$ grains is 3.7 $10^6$ yr, compared to the
cooling time scale of 2.1  $10^6$ yr.
Hence the grains can indeed survive the
shock long enough to be the major coolant.
The energy density of relativistic particles plus
the magnetic fields, derived from the minimum energy
assumption, is $\rm U_{\rm min} \sim 1.0\times 10^{-11}$
erg~cm$^{-3}$ (Xu et al.~\cite{xu03}). This is significantly lower than the
total energy density implied by the total energy flux,
so the magnetic field may be dynamically insignificant
for the shock. Recently, using Spitzer, 
Appleton et al. (\cite{apple06}) discovered strong mid-IR emission
lines of molecular hydrogen (and little else!) 
at position of
one of the radio emission peaks along the shock.
The derived H$_2$ luminosity is much higher than $\rm L_{\rm X}$
and only a factor of $\sim 3$ less than L$_{\rm dust}$.
This gives support to the idea (Rieke et al.~\cite{rie85}) that
the strong MIR H$_2$ lines often detected in luminous
IR galaxies are primarily excited by shock waves.

\section{Star formation}
The star formation activity in SQ is apparently very much influenced
by interactions.  The most spectacular star formation region in SQ
is the IGM starburst SQ-A, which is associated with a bright MIR
(15$\mu m$) source (Xu et al.~\cite{xu99}) just beyond the
northern tip of the shock front. 
The two velocity components of the starburst have same redshifts 
as those of the neutural gas, indicating that the starburst is occurring in
the pre-shock gas. The estimated star formation rate (SFR) is
1.45 M$_{\sun}$/yr, and the age is $\sim 10^7$ yr. 
Xu et al. (\cite{xu03}) advocated a scenario in which
the starburst is triggered due to squeezing of pre-existing giant-molecular
clouds (GMCs) by post-shock HI gas, as depicted in the model of Jog \& Solomon 
(\cite{js92}). The 'young tail' is also active in star
formation. Hunsberger et al. (\cite{hun96}) found 13 'tidal dwarf galaxy
candidates' along this tail.  A bright star formation region (SQ-B) is detected in
both H$\alpha$ (Arp~\cite{arp73a}) and MIR (Xu et al.~\cite{xu99}). 
Several discrete star
formation regions near the western tip of the tail are found in both
H$\alpha$ (S01; Mendes de Oliveira et al.~\cite{mdo04}) and
FUV (X05).  The H$\alpha$ images (Arp~\cite{arp73a}; Vilchez \&
Iglesias-P\'aramo~\cite{vi98} 1998; Plana et al.~\cite{plana99}; S01) 
reveal numerous huge
HII regions along several arms of NGC~7318b. Hunsberger et al. (\cite{hun96}),
Iglesias-P\'aramo \& Vilchez (\cite{iv01}), and Mendes de Oliveira et
al. (\cite{mdo01}) classified them as tidal dwarf galaxy candidates, though
S01 argued that the crossing time of NGC~7318b ($\sim 10^7$ yr) is
too short for any tidal effects. Investigations based on optical
colors (Schombert et al.~\cite{sch90}; Gallagher et al. \cite{gal01}) and UV colors
(X05) indicate several distinct epochs of star formation
that appear to trace the history of dynamic interactions in SQ. Using
the $H_\alpha$ luminosity without correction for internal
extinction, the results of Iglesias-P\'aramo \& Vilchez (\cite{iv01})
give a rather low net
SFR of $\sim 1$ M$_{\sun}$/yr for the whole group. 
The extinction corrected FUV (GALEX) luminosities, with internal
extinction constrained by the ISO MIR flux  and the HI
surface density, yield a net SFR of
6.7$\pm 0.6$ M$_{\sun}$/yr (X05). The SFR of the two Sbc galaxies, NGC~7318b
and 7319, is consistent with that of normal Sbc galaxies such as the
Milky Way, confirming the general conclusion that late type galaxies
in compact groups have their SFR indistinguishable from that
of their field counterparts (Sulentic \& De Mello Rabaca~\cite{sul94};
Moles et al.~\cite{mol94}; Iglesias-P\'aramo \& Vilchez~\cite{iv01}). 
This is in strong 
contrast with galaxy pairs where significant enhancement
of star formation is very common.

\section{Intra-Group Medium (IGM)}
\nid{\bf HI gas:}
First noticed by Allen \& Sullivan (\cite{allen80}), and confirmed by
later observations (Shostak et al.~\cite{sho84}; W02),
all of the HI in SQ is in the IGM outside galaxy disks.
W02 found the gas is located in five features:
(1) Arc-S (6641 km sec$^{-1}$, 2.5 10$^9$ M$_{\sun}$) associated with the old 
optical tail, (2) Arc-N (6604 km sec$^{-1}$, 4.0 10$^9$ M$_{\sun}$) associated
with the young tail, (3) NW-LV (6012 km sec$^{-1}$, 2.2 10$^9$ M$_{\sun}$) 
and (4) NW-HV 
(6647 km sec$^{-1}$, 0.9 10$^9$ M$_{\sun}$) 
both centered at the position of the IGM starburst
SQ-A, and (5) SW (5699 km sec$^{-1}$, 1.5 10$^9$ M$_{\sun}$) 
in the south of NGC~7318b.
These authors emphasized the uncertain origin of these 
HI features.
On the basis that SW and NW-LV features are spatially and kinematically
separated with each other, W02 argued against the scenario that
both features belong to NGC~7318b, each on the opposite
side of a rotation disk (M97). However, a comparison between the FUV
image (X05) and a new, higher resolution and higher
sensitivity VLA B-array HI map (Yun et al.~\cite{yun06}) shows that
SW and NW-LV indeed coincide very well with FUV arms belonging to
a very large FUV disk ($\sim 80$ kpc) centered on the nucleus
of NGC~7318b, lending support to their associations with
NGC~7318b.

\nid{\bf Molecular Gas:}
The early millimeter CO observations (Yun et al.~\cite{yun97}; 
Verdes-Montenegro et al.~\cite{vm98};
Leon et al.\cite{leon98})
 detected molecular gas only in the disk of NGC~7319, amounting
to 4.8 10$^{9}$ M$_{\sun}$ (Smith \& Struck~\cite{ss01}). This is normal for its 
size and luminosity (Lisenfeld et al.~\cite{lf02}). The molecular gas associated
with the IGM starburst SQ-A is first detected by 
Gao \& Xu (\cite{gao00}) using BIMA, later confirmed by single dish observations
of Smith \& Struck (\cite{ss01}) and Lisenfeld et al. (\cite{lf02}).
 The CO emission is
in two separate velocity systems
centered at 6000 km sec$^{-1}$ and 6600 km sec$^{-1}$, respectively, in
excellent agreement with the redshifts of the two HI features found
in the same region (NW-HV and NW-LV in W02).
It is interesting to note that, according to Lisenfeld et al. (\cite{lf02}),
there is more molecular gas (3.1 10$^9$ M$_{\sun}$) in SQ-A than in 
HI gas (2.5 10$^9$ M$_{\sun}$). 
Braine et al. (\cite{br01}) detected CO in SQ-B, a `tidal dwarf'  (TDF) candidate
bright in MIR (Xu et al.~\cite{xu99}), H$\alpha$ (Arp~\cite{arp73a}) and UV (X05).
The follow-up observations of Lisenfeld et al. (\cite{lf02}; \cite{lf04})
found 7 10$^{8}$ M$_{\sun}$ molecular gas. The ratio
M$_{H_2}$/M$_{HI} = 0.5$ is
consistent with the average of TDFs. From the close correspondence
between the CO, HI, H$\alpha$ and MIR emission, Lisenfeld et al. (\cite{lf04})
argued that the HI feature Arc-N (W02) is linked
to the young optical tail, therefore is very likely to be originated
from NGC~7319. A survey for CO
emission associated with other TDF candidates (Mendes de Oliveira et
al.~\cite{mdo01}) yielded no detection. CO emission associated with NGC~7318b
nucleus and with several other regions in the disk was reported
in Gao \& Xu (\cite{gao00}), Smith \& Struck (\cite{ss01}), Petitpas \& Taylor
(\cite{pt05}). Metallicity has been determined for SQ-A and SQ-B, 
in both cases it is slightly higher than solar (Xu et al.~\cite{xu03}; 
Lisenfeld et al.~\cite{lf04}), suggesting that the IGM gas is originated
in the inner part of a galaxy disk (or disks).

\nid{\bf Hot Gas (X-ray):}
SQ has been observed by almost every X-ray satellite, including
Einstein (Bahcall et al.~\cite{ba84}), Rosat (Sulentic et al.~\cite{sul95}; S01;
Pietsch et al.~\cite{pie97}), ASCA (Awaki et al.~\cite{aw97}),
Chandra (Trinchieri et al.~\cite{trin03}) and XMM (Trinchieri et al.~\cite{trin05}).
Trinchieri et al. (\cite{trin05}) divide
the X-ray emission in SQ into 4 major features:
(1) the shock, (2) NGC~7319 (Sy2), (3) HALO, (4) TAIL.
TAIL is associated with
hot diffuse IGM that cannot be ascribed to the shock.
It has  a size of 130 --- 150 kpc and is in
a region coinciding with the eastern end of the old optical tail,
indicating a link to the early tidal interactions.
HALO includes the diffuse emission in a
region surrounding the shock, excluding the shock itself and
emissions possibly related to individual galaxies.
These two diffuse features take 
$\sim 50\%$ of the total soft X-ray (0.5 --- 2.0 keV) luminosity
of SQ.

\nid{\bf Dust:}
SQ was detected, but barely resolved, 
by IRAS (Allam et al.~\cite{al96}; Yun et al.~\cite{yun97}). 
The higher resolution and more sensitive ISO
observations (Xu et al.~\cite{xu99}, \cite{xu03}; S01) found most
of the MIR (11.3 and 15$\mu m$) and FIR (60 and 100$\mu m$)
emission in the disks of NGC~7319 and of the foreground galaxy
NGC~7320. Two IGM starbursts SQ-A and SQ-B, both including several
previously detected HII regions, stand out in the 15$\mu m$ image due
to their strong MIR emission (Xu et al.~\cite{xu99}). There is evidence for
dust emission in the shock front, and the dust cooling is likely to be
the dominant cooling mechanism for the shock (Xu et al.~\cite{xu03};
Trinchieri et al.~\cite{trin05}).  The ISOPHOT maps at 60$\mu m$ and 100$\mu
m$ indicate that there might be diffuse dust emission in the IGM, 
but the signal is marginal. This has been followed up by
Spitzer observations at 70$\mu m$ and 160$\mu m$ (Xu et
al.\cite{xu06}). With much improved sensitivity and angular resolution, the
Spitzer 160$\mu m$ indeed detected significant diffuse dust emission outside
individual galaxies, amounting as much as more than 50\% of the total
L$_{160\mu m}$ of SQ.  The origin of this emission is still unclear.
There are two competing possibilities: (1) emission of collisionally
heated grains in the hot IGM; and (2) emission of radiative heated
grains in the cold neutral IGM.  The better correspondence between the
morphology of the 160$\mu m$ map and that of the X-ray map,
compared to that between 160$\mu m$ and HI, favors the possibility
(1).

\section{Sy2 Nucleus of NGC~7319}
The AGN was first recognized by van der Hulst and Rots (\cite{van81}) in the
VLA observations of SQ at 20cm continuum. They detected a bright compact
source with a jet-like extension, and suggested 
this might be related to those found in Seyfert galaxies. Huchra et al. (\cite{hu82}) 
obtained the optical spectrum which shows
clearly Sy2-type emission-line features. Based on the stellar velocity
dispersion, Woo \& Urray (\cite{woo02}) estimated
the black hole mass: M$_{\rm BH} = 10^{7.38}$ M$_{\sun}$.
Aoki et al. (\cite{aok99}) made VLA A-array radio continuum observations,
complimented with an archival HST optical image.
They found a chain of 3 radio sources,
interpreted as the nucleus and its two jets on opposite sides. 
Optical features are found in the HST image closely related to 
the radio jets, interpreted as gas compressed and excited by bow shocks 
driven into the ambient medium by the jets.
This is different from the so-called extended emission line regions (EELR)
which are supposedly excited by the AGN radiation (Aoki et al.~\cite{aok96}).
Even higher resolution (0.16'') radio continuum observations at
1658 MHz, using MERLIN, were reported by Xanthorouplos et al. (\cite{xan04}).
They compared the data with an HST/ACS U-band (F330W) image, and found
extended UV emission around the nucleus and the northern jet. 
They argued that this indicates star formation triggered by
the jet/ISM interaction. By assuming the diffuse radio emission
outside the compact sources is due to the star formation, 
they estimated that the SFR in the circum-nuclear region 
is 8.4 M$_{\sun}$. However, all of the optical spectra of emission
line regions in the circum-nuclear region show Seyfert or
LINER line ratios and none is HII region-like (Aoki et al.~\cite{aok96}), 
indicating that most of the radio and UV radiation is related to the AGN and/or
the shocks. Comparing the UV and MIR observations, X05 found the AGN and
the surrounding region is highly obscured, with an FUV extinction of
$A_{\rm FUV} = 5.4$ mag, consistent with the Sy2 classification.

\section{Summary Remarks}
The fascination of SQ is due to its
rich history full of interactions, which we are just starting
to unveil. Although the multi-wavelength observations in the literature
have provided clues on how the different events may be related to
each other, a clear-cut picture is yet to emerge. Many interpretations
are based on dynamic simulations for binary galaxies, which may not
be appropriate for multi-galaxy collisions such as SQ.
Dedicated simulations for SQ are certainly desired.
Such simulations will provide insights on how compact groups such
as SQ survive merging, at the same time turn late-type galaxies into
early types without significantly increasing the SFR.
Investigations on the 
difference between multi-galaxy collisions and two-galaxy collisions
will shed light on how galaxy interactions may have contributed to
the evolution of galaxies and galactic structures.


\label{lastpage}

\end{document}